\documentclass[useAMS, usenatbib]{mnras}
\usepackage{amsmath,color}
\usepackage{slantsc}
\usepackage{graphicx}

\begin{document}
\bibliographystyle{mnras} \title[]{Globular clusters {\it vs} dark
  matter haloes in strong lensing observations }

\author[Qiuhan He et. al]
       {\parbox[t]{\textwidth}{
	       Qiuhan He$^{1,2}$, Ran Li$^{1,2}$\thanks{E-mail:ranl@bao.ac.cn}, Sungsoon Lim$^{3,4}$, Carlos S. Frenk$^{5}$, Shaun Cole$^{5}$, Eric W. Peng$^{3,4}$,  Qiao Wang$^{1,2}$
       }
        \vspace*{3pt} \\
  $^{1}$Key laboratory for Computational Astrophysics, National Astronomical Observatories, Chinese Academy of Sciences, Beijing, 100012, China\\
    $^{2}$University of Chinese Academy of Sciences, 19 A Yuquan Rd, Shijingshan District, Beijing, 100049, China\\
    $^{3}$Department of Astronomy, Peking University, Beijing, 100871, China\\
    $^{4}$Kavli Institute for Astronomy and Astrophysics, Peking University, Beijing, 100871, China\\
    $^{5}$Institute for Computational Cosmology, Department of Physics, University of Durham, South Road, Durham, DH1 3LE\\
          }

\maketitle

\begin{abstract}
  Small distortions in the images of Einstein rings or giant arcs
  offer the exciting prospect of detecting dark matter haloes or
  subhaloes of mass below $10^9$M$_{\odot}$, most of which are too
  small to have made a visible galaxy. A very large number of such
  haloes are predicted to exist in the cold dark matter model of
  cosmogony; in contrast other models, such as warm dark matter,
  predict no haloes below a mass of this order which depends on the
  properties of the warm dark matter particle. Attempting to detect
  these small perturbers could therefore discriminate between
  different kinds of dark matter particles, and even rule out the cold
  dark matter model altogether. Globular clusters in the lens galaxy
  also induce distortions in the image which could, in principle,
  contaminate the test. Here, we investigate the population of
  globular clusters in six early type galaxies in the Virgo cluster.
  We find that the number density of globular clusters of mass
  $\sim10^6$M$_{\odot}$ is comparable to that of the dark matter
  perturbers (including subhaloes in the lens and haloes along the
  line-of-sight). We show that the very different degrees of mass
  concentration in globular clusters and dark matter haloes result in
  different lensing distortions. These are detectable with
  milli-arcsecond resolution imaging which can distinguish between
  globular cluster and dark matter halo signals.

\end{abstract}

\section{Introduction}
Perhaps the most fundamental prediction of cosmogonic models in which
the dark matter consists of cold collisionless particles (CDM), such
as the current cosmological paradigm, $\Lambda$CDM, is the existence
of a very large number of dark matter haloes with masses extending
well below the masses of even the faintest galaxies known (for
example, down to an Earth mass for a few GeV weakly interacting
particle
\citet{Green2005})\citep{Diemand2007,Springel2008,Frenk2012}. This
property distinguishes CDM models from models in which the dark matter
is a warm particle (WDM), such a sterile neutrino
\citep{Boyarski2009}. In this case, particle free streaming in the
early universe induces a low-mass cutoff in the power spectrum of
density perturbations. As a result, few dark matter haloes of mass less
than about $10^8$M$_{\odot}$ (depending on the particle properties,
\citep[see][]{Lovell2016,Bose2016}) ever form. Searching for haloes
smaller than about $10^8$M$_{\odot}$ provides a clearcut way to
distinguish between these dark matter candidates. Finding a dark
matter halo of mass, say, $10^7$M$_{\odot} $ would rule out most WDM
models of interest. Conversely, failing to find such haloes would
conclusive rule CDM models, including $\Lambda$CDM.

A major difficulty in applying this test is that the haloes of interest
are too small ever to make a galaxy and are thus completely dark
\citep[see][and references therein]{Sawala2016}. For this reason
attempts to test CDM or distinguish it from WDM using the abundance of
the more massive haloes that do make a galaxy are misguided: many
studies dating back to semi-analytic models in the early 2000s
\citep{Bullock2000,Benson2002,Somerville2002} and, more recently, using
hydrodynamic simulations
\citep[e.g.]{Okamoto2009,Maccio2006,Sawala2016,Wetzel2016} have
clearly demonstrated that physically-based $\Lambda$CDM models predict
the correct number of faint galaxies, including satellites of the
Milky Way. By contrast, the observed abundance of satellites can be
used to rule out regions of the WDM parameter space
\cite{Kennedy2014,Lovell2016}.

A conclusive test of CDM and many other particle dark matter
candidates requires counting dark, small-mass haloes. Fortunately these
are detectable through gravitational lensing, specifically through the
distortions they induce in strongly lensed systems that produce
Einstein rings or giant arcs
\citep{Koopmans2005,Vegetti2009a,Vegetti2009b,Vegetti2012,Hezaveh2016,Li2016}.
Until recently, it was thought that the distortions would come from
substructures inside the lens. However, \cite{Li2017} have recently
shown that the distortions in CDM and WDM models are dominated by
intervening haloes projected onto the Einstein ring or strong lens
rather than by subhaloes in the lens. This is a fortunate feature
because it makes the test of particle candidates particularly clean:
small-mass intervening haloes are not affected in any way by baryon
effects, unlike subhaloes some of which can be destroyed by interaction
with the central galaxy \citep{Sawala2017,Garrison-Kimmel2017}.

For the lensing test to be effective, it is essential to be able to
detect haloes of small mass \citep{Li2016}. The detection limit of a
particular observation depends on the resolution of the imaging
data. High resolution imaging with the Hubble Space Telescope (HST) can
reveal haloes down to at least $\sim10^9$M$_{\odot}$
\citep{Vegetti2010}, while adaptive optics on the Keck telescope can
reveal haloes down to $\sim10^7$M$_{\odot}$ \citep{Vegetti2012}.
Simulations show that even smaller haloes, down to $10^6$M$_{\odot}$,
can be detected in radio-selected strong lenses \citep{McKean2015}.

\cite{Li2016} recently investigated the sample size and detection
limit required to distinguish $\Lambda$CDM from the particularly
interesting WDM model in which the particles are 7~keV sterile
neutrinos whose decay could explain the 3.5 keV X-ray line recently
detected in galaxies and clusters \citep{Boyarski2014,Bulbul2014}. In this
case, the cutoff in the halo mass function occurs at a mass of $\sim10^8$M$_{\odot}$ \citep{Bose2016}. With a detection limit of $10^7$M$_{\odot}$, $20\sim100$ lenses would suffice to distinguish even the
coldest 7 kev sterile neutrino model allowed by the X-ray data from
CDM. \citet{Li2016} also showed that the constraining power increases
rapidly with improvements in the detection limit.  If VLBI imaging
could really reach a detection limit of $10^6$M$_{\odot}$, a handful of
lens systems would already be enough conclusively to rule out all
7~keV sterile neutrinos, if a signal is found, or CDM if it is not.

A potential complication arises because dark matter haloes and subhaloes
are not the only objects that can perturb an Einstein ring. Globular
clusters in the lens galaxy could also be a source of
perturbations. Studies of nearby galaxies show that the mass function
of globular clusters has a Gaussian distribution peaking at about
$10^5$M$_{\odot}$ \citep{Jordan2007}. In the Milky Way, the largest
globular cluster is $\omega$ Centauri, which has a mass of
$4.05\pm0.1\times10^6$M$_{\odot}$ \citep{DSouza2013}. The most massive
globular clusters cannot generate lensing signals strong enough to be
detected in optical lenses at HST imaging resolution, but
they could be detect in radio lens systems with VLBI imaging resolution.

In this paper we estimate the number density of globular clusters near
the Einstein radius of typical early type galaxies. Using a catalogue
of globular clusters in the Virgo cluster \citep{Eric2008, Jordan2009}, we
estimate the strong lensing distortions induced by globular clusters
and compare them to the distortions induced by dark matter haloes of
the same mass.

The paper is organized as follows. We provide a brief description of
the globular cluster catalogue in Section~\ref{sec:gc}. In
Section~\ref{sec:number} we calculate the number density of globular
clusters in the Einstein ring region.  In Section~\ref{sec:raytracing}
we calculate the differences in the lensing effects of globular
clusters and NFW haloes. We summarize our conclusions in Section
\ref{sec:sum}

\section{The globular cluster catalogue}
\label{sec:gc}

To investigate the distortions induced by globular clusters in strong
lensing systems we first estimate the total number of globular
clusters expected in the lens galaxy using the catalogue of globular
clusters around galaxies in the Virgo cluster compiled by
\citet{Eric2008} and \citet{Jordan2009}.

\begin{figure}
\centering
\includegraphics[width=0.5\textwidth]{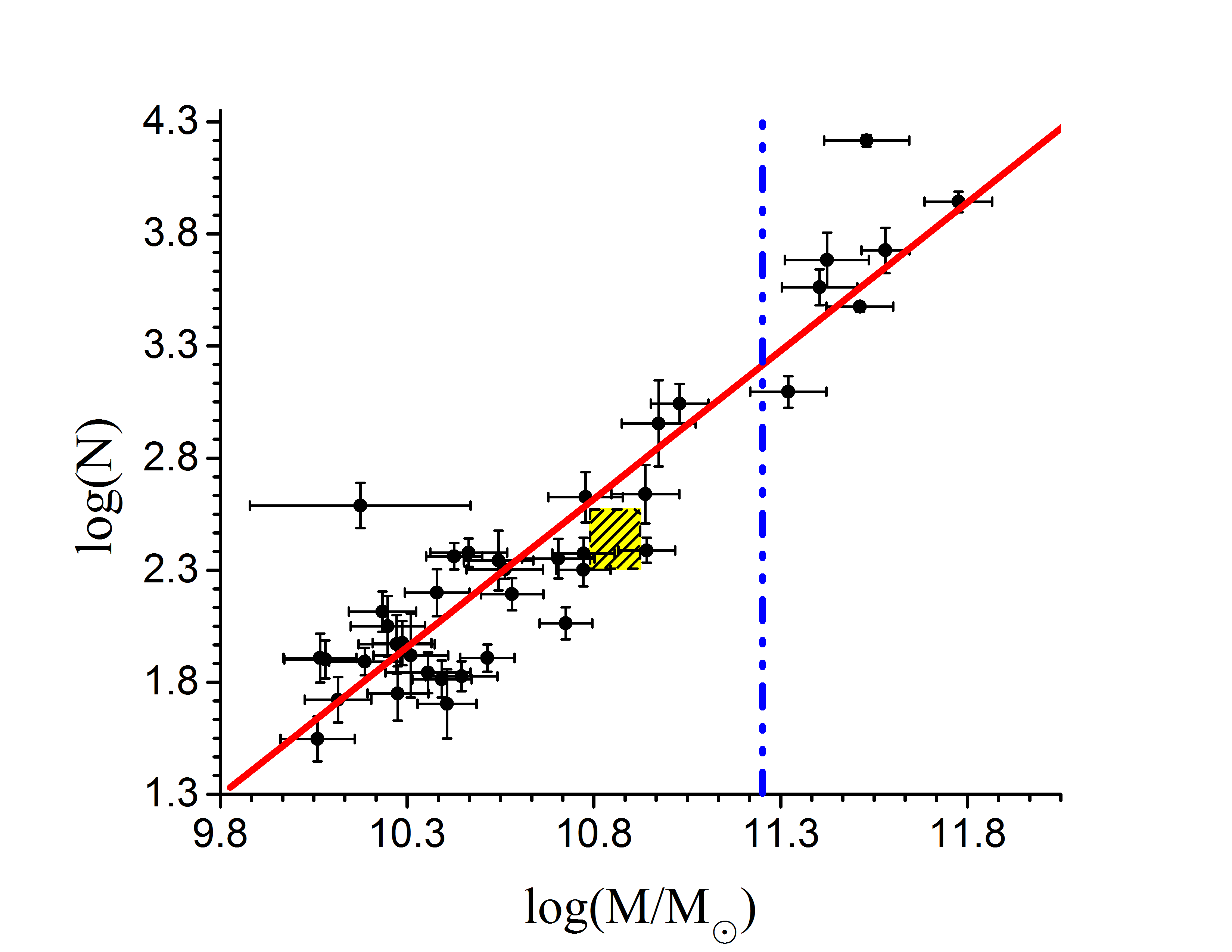}
\caption{The relation between the number of globular clusters, $N$ and
  the host galaxy stellar mass. The rectangle marks the location of
  the Milky Way in this plane. Galaxies to the right of the vertical
  dashed line have similar masses to the galaxies in the SLACS sample
  \citep{Bolton2008}.}\label{fig:1}
\end{figure}

Fig.\ref{fig:1} shows that the total number of globular clusters in
this catalogue increases as a function of galaxy stellar mass. A
typical lens galaxy in the SLACS sample \citep{Bolton2008} has a
stellar mass of a few $10^{11}$M$_{\odot}$. According to
Fig. \ref{fig:1}, these galaxies possess more than 5000 globular
clusters each.  The total number of globular clusters increases as
$M^{1.45}$, where $M$ is the stellar mass of the globular
cluster. The number of globular clusters is known to be proportional to stellar mass, and may in fact be linear with total galaxy halo mass \citep{Blakeslee1997,Eric2008,Hudson2014}.

\section{The number density of globular clusters}
\label{sec:number}

\subsection{Projected number density profile}
The projected number density of globular clusters as a function of
projected radius, $R$, is well described by  a S\'{e}rsic
\citep{Sersic1968} profile:
\begin{equation}\label{ser}
\Sigma(R) = \Sigma_{e}\exp\left[-b_{n}\left(\left(\frac{R}{R_{e}}\right)^{1/n}-1\right)\right] \,,
\end{equation}
where
\begin{displaymath}
b_{n} = 2n - \frac{1}{3} \,,
\end{displaymath}
with three model parameters, $\Sigma_{e}, R_{e}$ and $n$. In this
paper,we use the best-fit model parameters for globular clusters in
individual Virgo galaxies derived by Lim et al. (2017 in prep.)

\subsection{Mass function}

In the Einstein ring model only structures whose mass exceeds a
certain threshold, $M_{\rm th}$, can be detected. If the globular
cluster mass function is independent of distance from the centre of
the galaxy, then the surface number density of globular clusters
within the projected Einstein radius, $R_{\rm Ein}$, more massive than
$M_{\rm th}$ can be written as:
\begin{equation}
\Sigma'(>M_{\rm th} , R_{\rm Ein})=\Sigma(R_{\rm Ein})  \int_{M_{\rm th}}^{\infty}\frac{dn}{dM}(M)dM \,,
\end{equation}
 where $dn/dM$ is the normalized globular cluster mass function
(GCMF), which may be inferred from the globular cluster luminosity
function (GCLF) by assuming a (constant) mass-to-light ratio,
$\Upsilon$. We choose widely used Gaussian GCLFs described by \citet{Jordan2007}:
\begin{equation}
\frac{dN}{dm}=\frac{1}{\sqrt{2\pi}}\exp\left[-\frac{(m-\mu_m)^{2}}{2\sigma_m^{2}}\right],
\end{equation}
where $m$ is the absolute magnitude of a globular cluster, $\mu_m$ is
the mean globular cluster magnitude, $\mu_m=<m>$ and the dispersion
$\sigma_m= <(m - \mu_m)^2>^{1/2}$. The form
of the GCMF may be written as,
\begin{equation}\label{equ:massf}
\frac{dN}{dM}=\frac{1}{\textrm{ln(10)}M}\frac{1}{\sqrt{2\pi\sigma_{M}}}\exp\left[-\frac{(\textrm{log}M-\langle \textrm{log}M \rangle)}{2\sigma_{M}^{2}}\right],
\end{equation}
where the mass and magnitude of a globular cluster are related by
\begin{equation}\label{equ:Mmr}
m = C - 2.5\textrm{log}M \,,
\end{equation}
and we have
\begin{equation}
\sigma_{M} = \frac{\sigma_{m}}{2.5},
\end{equation}
and
\begin{equation}\label{equ:m}
\begin{aligned}
\langle \textrm{log}M \rangle&= \int (C-0.4m)\frac{1}{\sqrt{2\pi}}\exp\left[-\frac{(m-\mu_{m})^{2}}{2\sigma_{m}^{2}}\right]dm\\
	 &= C - 0.4\mu_{m};
\end{aligned}
\end{equation}
the constant, $C$, is related to $\Upsilon=M/L$ through
\begin{equation}
C=0.4m_{\odot}+\log{\Upsilon} \,.
\end{equation}
where $m_{\rm \odot}$ is the absolute magnitude of the Sun.

\subsection{Estimate for six massive galaxies in the Virgo cluster}

Combining the mass function and number density profile we can estimate
the number of globular clusters in a given mass range projected onto a
specific annulus.  We estimate the project number density of globular
clusters in six massive galaxies in the globular cluster
catalogue of the Virgo cluster. The stellar masses of these galaxies
are comparable to those of the lens galaxies in the SLACS sample. We
list the mass of these six galaxies in Table \ref{tab:1}.

We use the values of $\mu_{m}$ and $\Upsilon_{m}$ derived by
\citet{Jordan2007} to calculate the GCLF of galaxies in the Virgo
cluster in the $z$-band. \cite{Jordan2007} showed that $\Upsilon$ is
nearly constant with a value of around 1.5 in this band, so we fix
$\Upsilon=1.5$ throughout this paper.

\begin{figure}
\centering
\includegraphics[width=0.5\textwidth]{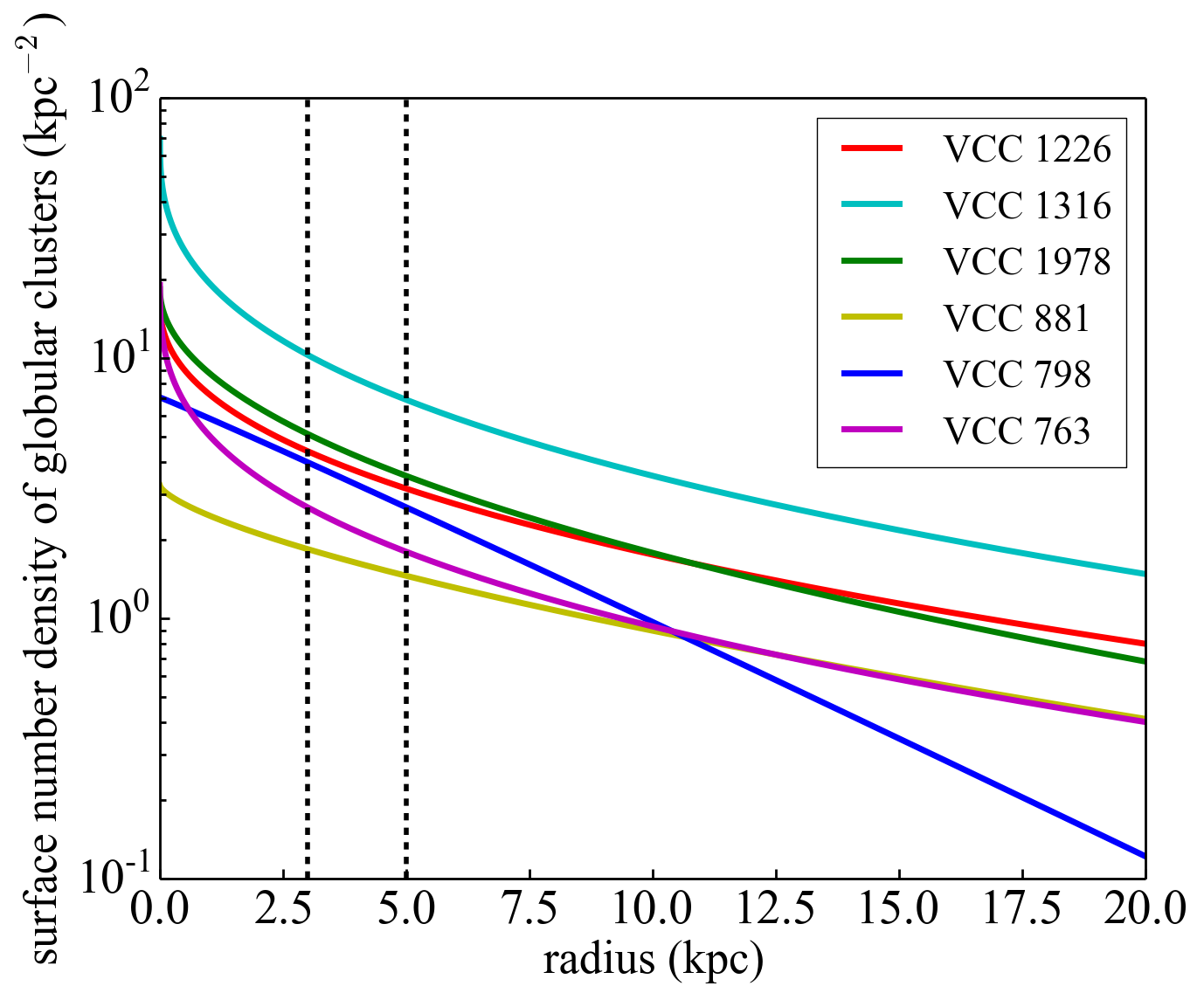}
\caption{Globular cluster number density profiles of six massive
  galaxies in the Virgo cluster.  The region between the two dashed
  vertical lines is the range of Einstein radii of typical SLACS
  lenses. }
\label{fig:2}
\end{figure}

The surface number density profiles of globular clusters in these six
galaxies are shown in Fig.~\ref{fig:2}. With the exception of VCC 798,
all other five galaxies have globular cluster surface density profiles
of similar form.

We calculate the number density of globular clusters in the annulus
between 3 kpc and 5 kpc from the galaxy centre, which is about
the Einstein radius of a typical lens in the SLACS survey. The
lenses in this survey have a mean velocity dispersion of 275 km s$^{-1}$ and
a mean redshift of $z=0.2$.  The results are listed in
Table~\ref{tab:2}.

Using $\Lambda$CDM N-body simulations \citet{Xu2015} derived the
surface number density of subhaloes in a $\Lambda$CDM universe and
found that it to about 1.5 arcsec$^{-2}$ in the strong lensing region of
a halo of $10^{13} h^{-1}$M$_{\odot}$. As mentioned in the Introduction,
in a recent paper, \cite{Li2017} showed that the dominant contribution
to perturbations in Einstein ring images comes from haloes along the
line-of-sight rather than from subhaloes.  The number of perturbing
line-of-sight haloes can be 3 times larger than the number of subhaloes.  From Table \ref{tab:2}, the number density of globular
clusters more massive than $10^6$M$_{\odot}$ is about $1.5$ to 9 $\rm
arcsec^{-2}$, which is comparable to the total number density of dark
matter haloes. The mass function of globular clusters drops sharply
above $10^6$M$_{\odot}$ so, for perturber masses above
$10^7$M$_{\odot}$, the contribution of globular clusters is negligible
compared with that of dark CDM haloes. Thus, our previous conclusions
regarding perturbers of mass $\sim10^9$M$_{\odot}$ detectable with HST
imaging remain unchanged.  However, future observations of
radio-selected lenses with VLBI high-resolution imaging will be
strongly affected by the presence of globular clusters.

\begin{table}
  \caption{Galaxy stellar mass and NGC number in
    the Virgo Cluster Catalogue (VCC).}\label{tab:1}
\centering
\begin{tabular}{cc}
\hline
Name & Stellar mass ($10^{11}$M$_{\odot}$)\\
VCC 1226/NGC 4472 & $5.32\pm1.10$\\
VCC 1316/NGC 4486 & $3.02\pm0.79$\\
VCC 1978/NGC 4649 & $3.39\pm0.50$\\
VCC 881/NGC 4406 & $2.90\pm0.60$\\
VCC 798/NGC 4382 & $1.87\pm0.44$\\
VCC 763/NGC 4374 & $2.36\pm0.61$\\
\hline
\end{tabular}
\end{table}
\begin{table*}
\caption{Model predictions for the surface number density of globular
  clusters around the Einstein rings of 6 massive early type galaxies
  in the Virgo cluster. }\label{tab:2}
\centering
\begin{tabular}{cccc}
\hline
Name &$ >10^{5}$M$_{\odot}$ &$>10^{6}$M$_{\odot}$ &$>10^{7}$M$_{\odot}$ \\
VCC 1226/NGC 4472&$29.34\pm8.62$ &$4.00\pm1.29$ &$0.0298\pm0.0167$ \\
VCC 1316/NGC 4486&$69.79\pm12.85$ &$8.94\pm1.79$ &$0.0431\pm0.0155$ \\
VCC 1978/NGC 4649&$33.37\pm14.88$ &$4.45\pm2.06$ &$0.0316\pm0.0202$ \\
VCC 881/NGC 4406 &$13.73\pm2.39$ &$1.78\pm0.45$ &$0.0085\pm 0.0064$ \\
VCC 798/NGC 4382 &$27.04\pm8.92$ &$3.12\pm1.17$ &$0.0112\pm0.0090$ \\
VCC 763/NGC 4374 &$19.00\pm12.90$ &$2.26\pm1.57$ &$0.0062\pm0.0057$ \\

\hline
the unit is $\textrm{arcsec}^{-2}$
\end{tabular}
\end{table*}

\section{The lensing effect of globular clusters}
\label{sec:raytracing}

Dark matter haloes and globular clusters have very different density
profiles and thus generate different lensing signals. In this section
we compare the distortions induced on the strong lensing images by
these two types of perturbers.

We assume that the main lens lies at $z_l=0.2$ and we model its
density profile with a singular isothermal sphere (SIS) of velocity
dispersion, $\sigma_v=275$ $\rm km$ s$^{-1}$. A globular cluster or a
dark matter subhalo is then placed at the Einstein radius. A source
galaxy whose surface brightness profile is taken to be a Gaussian of
dispersion, $\sigma_{\rm source}=0.02^{\prime \prime}$, is assumed to
be located at redshift $z_s=0.7$. Using a ray-tracing code we generate
a lensed image on a plane of 8000$\times$8000 pixels, each of size 0.7
milli-arcsecond.

We assume the dark matter subhalo follows a
density profile given by the NFW formula \citep{Navarro1996,Navarro1997}:
\begin{equation}
\rho (x)= \frac{\rho_{s}}{x(1+x)^2},
\end{equation}
where $x = r/r_s$ and $r_s$ is the scale
parameter. The concentration parameter, $c$, is defined as
\begin{equation}
r_s = r_{200}/c,
\end{equation}
where $r_{200}$ is the radius within which the average interior
density equals 200 times the critical density. We adopt the halo
mass-concentration relation derived by \citet{Neto2007} from the
Millennium simulation \citep{Springel2005}:
\begin{equation}
c=\frac{4.67}{(1+z)}\left(\frac{M_{200}}{10^{14}h^{-1}\text{M}_{\odot}}\right)^{-0.11} \,.
\end{equation}
The density profile of the globular cluster is represented with a King
model \citep{King1966}. Its density, $\rho(r)$, can be obtained by
solving the following equations:
\begin{subequations}
\begin{align}
&\frac{d^{2}W}{dR^{2}} + \frac{2}{R}\frac{dW}{dR} = -9\frac{\rho}{\rho_{0}}\label{equ:14a}\\
&\rho = \frac{9}{\rho_0}\exp\left[W-W_{0}\right]\int_{0}^{W}\eta^{-3/2}e^{-W}d\eta\label{equ:14b}\\
&R = r/r_{0}\label{equ:14c}\\
&\lim_{R\to 0} \frac{2}{R}\frac{dW}{dR} = -6,
\label{equ:14d}
\end{align}
\end{subequations}
where $r_{0}$ is a scale parameter and $W_{0}$ and Eq.~\ref{equ:14d}
are the initial conditions for Eq.~\ref{equ:14a}. The King model has
three parameters: $W_{0}$, $\rho_{0}$ and $M_{gc}$, the mass of the
globular cluster.

\begin{figure}
\centering
\includegraphics[width=0.5\textwidth]{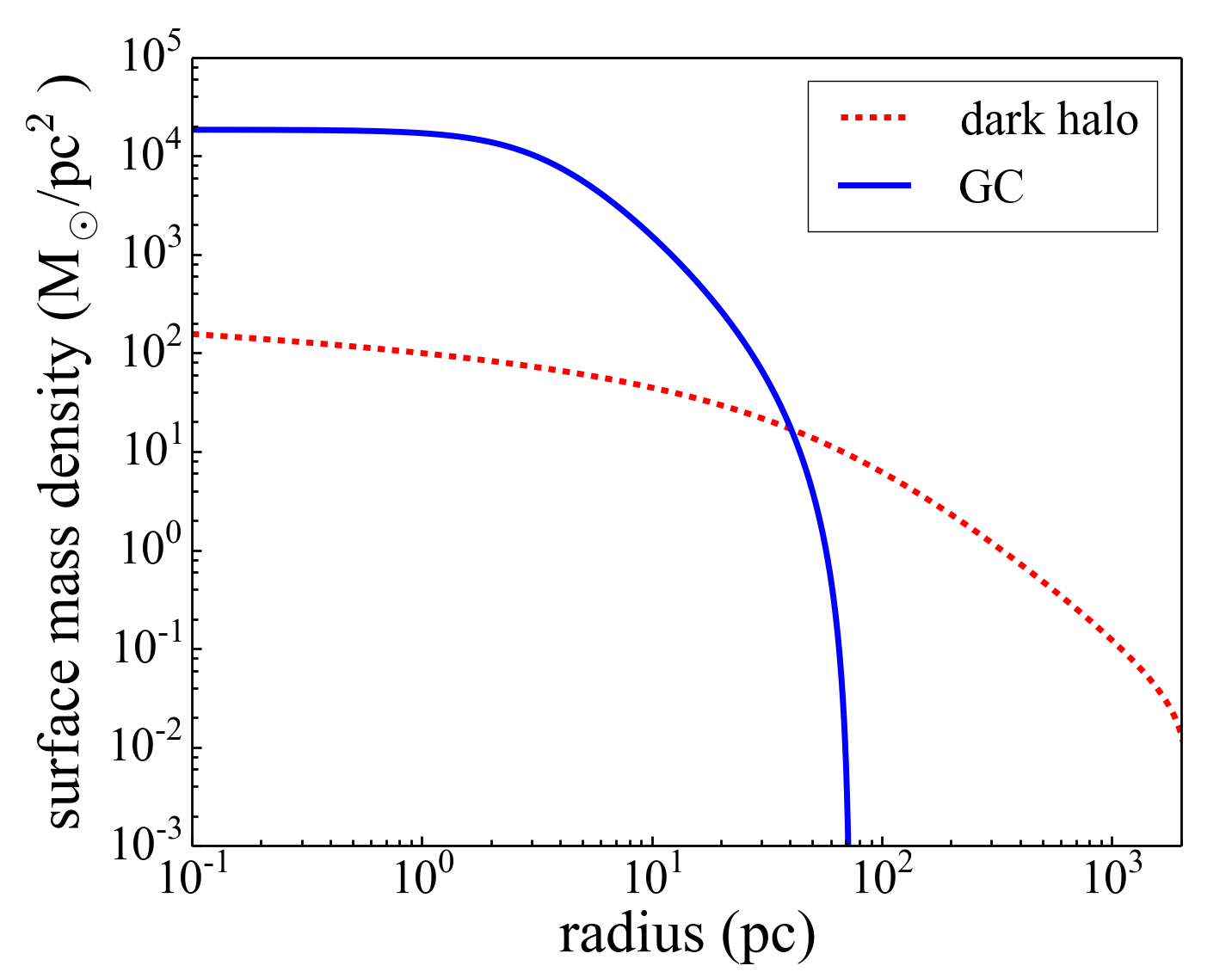}
\caption{The surface density profile of a globular cluster (solid
  line) and a dark matter halo (dotted line) of the same mass. The
  unit of length is a parsec and the unit of surface mass density is M$_{\odot}$pc$^{-2}$} \label{fig:massd}
\end{figure}

To compare the surface mass density profiles of our two kinds of
perturbers, we consider the most massive globular cluster in the
compilation of \cite{Mc2005}, NGC 5139, whose King model parameters
are $M_{gc}=10^{6.37}$M$_{\odot}$, $W_{0}=6.2$ and $\rho_{0} =
10^{3.43}$M$_{\odot} {\rm pc}^{-3}$. In Fig.~\ref{fig:massd}, we compare
the surface density profile of NGC~5139 with the profile of a NFW halo
of the same mass. It is clear that the globular cluster is much more
concentrated than the NFW subhalo. The radius of the globular cluster
is 72 pc, while the radius of the much more extended dark halo is
$r_{200} \sim 2000$ pc.

\begin{figure*}
\centering
\includegraphics[width=1\textwidth]{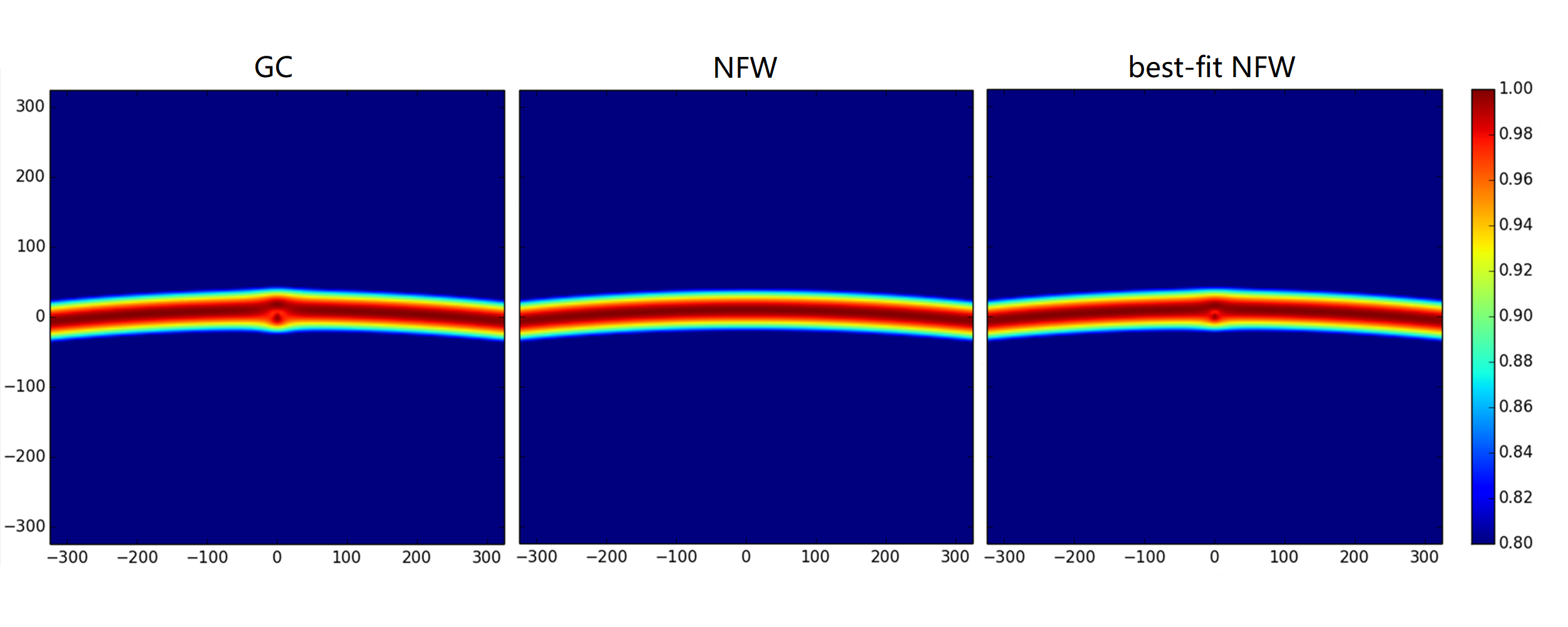}
\caption{Image of a section of an Einstein ring. The left-hand panel
  illusrates the lensing effect of a globular cluster of mass
  $10^{6.37}$M$_{\odot}$, while the central panel illustrates the
  lensing effect of an NFW halo of the same mass. The right-hand panel
  shows the result of attempting to fit the distortion in the
  left-hand panel, assuming incorrectly that it is due to an NFW
  halo. The best-fit NFW halo has mass of $10^{6.54}$M$_{\odot}$ and
  concentration, $c=10^{4.45}$. The axes, showing corresponding
  physical size in the lens plane, are in unit of pc.}

 \label{fig:lens}
\end{figure*}

In Fig.\ref{fig:lens} we illustrate the distortions induced on a
lensed image by the different kinds of perturber. Each panel shows
a section of an Einstein ring near the projected position of the perturber. The
presence of the perturber distorts the image around it. For the
globular cluster, the induced distortion changes the surface brightness
around its position by about 2\%, while for the NFW halo of the same mass,
no distortion is visible on the plot.

The right-hand panel of Fig.\ref{fig:lens} shows the result of
force-fitting the perturbation induced by the globular cluster
assuming, incorrectly, that it is due to an NFW halo.  The best-fit mass
of the NFW model is $10^{6.54}$M$_{\odot}$, which is about 1.5 times the input
value for the globular cluster, and the best-fit concentration, $c$,
is $10^{4.45}$, which is about 1200 times the value predicted for a
cold dark matter halo of that mass \citep{Neto2007}. For six galaxies listed above, the average number density for globular clusters with mass larger than $10^{6.37}$M$_\odot$ is 0.88 arcsec$^{-2}$, which is larger than that of dark matter halos in the lens plane with mass larger than $10^{6.54}$M$_\odot$ showed in \cite{Li2017}, but this will not have great effects on the detection of dark matter halos, since we note that the NFW profile actually cannot fit the image perturbed by the globular cluster. Although the
amplitude of the variation in brightness caused by this ultra-dense
NFW halo is comparable to that caused by the globular cluster, the
differences in the images are still visible by eye.  This result
demonstrates that although a globular cluster can generate a lensing
signal that is strong enough to be detected with VLBI imaging, proper
modelling can clearly distinguish between the distortions produced by
globular clusters and by dark matter haloes.

\section{Discussion and Summary}
\label{sec:sum}

Small distortions of images of Einstein rings or giant arcs offer the
exciting prospect of detecting dark matter haloes or subhaloes too small
to have made a visible galaxy. Since a fundamental property of the
cold dark matter model of cosmogony, which distinguishes it from other
possibilities such as warm dark matter, is the existence of a very
large number of such small haloes, detecting them could discriminate
between different kinds of dark matter particles, and even rule out the
cold dark matter model altogether.

A possible source of contamination of the signal are globular
clusters which can also distort a strong lensed image. These
distortions are detectable with milli-arcsecond resolution imaging.
In this paper we have calculated the lensing effect of globular
clusters and compared it to the lensing effect of intervening dark
matter haloes.

We selected six early type galaxies in the Virgo cluster of stellar
mass $\sim 10^{11}$M$_{\odot}$, similar to the mass of strong lens
galaxies in the SLAC survey. The number density of globular clusters
of mass $\sim 10^6$M$_{\odot}$ is between 1.5 to 9 arcsec$^{-2}$,
which is comparable to the number of dark matter perturbers, when both
subhaloes and line-of-sight haloes in the same mass range are
counted. These globular clusters are not massive enough to be detected through their lensing effects with images taken by Hubble Space Telescope at the typical lens redshift, but can be detected with VLBI imaging of milli-arsecond resolution.

We used a ray-tracing method to compare the distortions of the image
of an Einstein ring or giant arc induced by a globular
cluster and by an NFW halo. We find that the globular cluster produces
a much stronger lensing signal than an NFW halo of the same mass,
because the density profile of a globular cluster is much more
centrally concentrated than that of a dark matter halo.

Imaging at milli-arsecond resolution can therefore distinguish a
globular cluster from a dark matter halo. If the NFW density profile
is used to model the distortion caused by a globular cluster a poor
fit is obtained and the inferred concentration parameter is orders of
magnitute higher than the  concentration of a real NFW halo of the same
mass. We conclude that globular clusters will not compromise efforts
to measure the abundance of low mass dark matter haloes and subhaloes,
but their detection would be a byproduct of efforts to constrain the
identity of the dark matter from the strong lensing test.

\section*{ACKNOWLEDGEMENTS}
RL acknowledges NSFC grant (Nos.11303033, 11511130054,11333001),
support from the Youth Innovation Promotion Association of CAS and
Nebula Talent Program of NAOC and Newton Mobility award.  Qiuhan is
supported by undergraduate research program of CAS. EWP and SL acknowledge support from the National Natural Science Foudation of China under grant No. 11573002. This work was
supported in part by STFC Consolidated Grant ST/L00075X/1 to Durham
and by ERC Advanced Investigator grant, COSMIWAY.  This work used the
DiRAC Data Centric system at Durham University, operated by the
Institute for Computational Cosmology on behalf of the STFC DiRAC HPC
Facility (www.dirac.ac.uk). This equipment was funded by BIS National
E-infrastructure capital grant ST/K00042X/1, STFC capital grants
ST/H008519/1 and ST/K00087X/1, STFC DiRAC Operations grant
ST/K003267/1 and Durham University. DiRAC is part of the National
E-Infrastructure.

\bibliography{ref}

\end{document}